\RequirePackage{arydshln}
\documentclass[aps,twocolumn,nofootinbib,superscriptaddress,preprintnumbers,pra,10pt,floatfix]{revtex4-1}

\usepackage[utf8]{inputenc}

\usepackage{float}
\usepackage{amsmath,amssymb}
\usepackage{dsfont} 
\usepackage{hyperref}
\usepackage{graphicx}
\usepackage{enumitem}
\usepackage{mathtools}
\usepackage{bbold}
\usepackage{multirow}
\usepackage{ytableau}
\usepackage{youngtab}
\usepackage{braket}
\usepackage{soul}
\usepackage{cancel}
\usepackage{slashed}
\usepackage[normalem]{ulem}

\usepackage{lipsum}

\usepackage{colortbl} 

\definecolor{Gray}{gray}{0.95}
\definecolor{RGray}{gray}{0.90}
\definecolor{CGray}{gray}{0.92}

\usepackage{arydshln}

\makeatletter
\g@addto@macro\bfseries{\boldmath}
\makeatother

\makeatletter
\renewcommand\paragraph{\@startsection{paragraph}{4}{\z@}%
                                    {3.25ex \@plus1ex \@minus.2ex}%
                                    {-1em}%
                                    {\normalfont\normalsize\bfseries}}
\makeatother

\allowdisplaybreaks
\begin{document}

\preprint{}
\preprint{}

\title{ Probing Lepton Flavor Violation in Meson Decays
with LHC Data}

\author{S.~Descotes-Genon}
\email{sebastien.descotes-genon@ijclab.in2p3.fr}
\affiliation{Universit\'e Paris-Saclay, CNRS/IN2P3, IJCLab, 91405 Orsay, France}
\author{D.~A.~Faroughy}
\email{darius.faroughy@rutgers.edu}
\affiliation{NHETC, Department of Physics and Astronomy, Rutgers University, Piscataway, NJ 08854, USA}
\author{I.~Plakias}
\email{ioannis.plakias@ijclab.in2p3.fr}
\affiliation{Universit\'e Paris-Saclay, CNRS/IN2P3, IJCLab, 91405 Orsay, France}
\author{O.~Sumensari}
\email{olcyr.sumensari@ijclab.in2p3.fr}
\affiliation{Universit\'e Paris-Saclay, CNRS/IN2P3, IJCLab, 91405 Orsay, France}

\begin{abstract}
\vspace{3mm}
In this letter, we use LHC data from the Drell-Yan processes $pp\to\ell_i\ell_j$ (with $i\neq j$) to derive model-independent upper limits on lepton-flavor-violating meson decays. Our analysis is based on an Effective Field Theory (EFT) approach and it does not require a specific assumption regarding the basis of effective operators. We find that current LHC data (140~$\mathrm{fb}^{-1}$) already provides competitive limits on $\mathcal{B}(B\to \pi e \tau)$ and $\mathcal{B}(B\to \pi \mu \tau)$ with respect to the ones obtained through experimental searches at the $B$-factories. Moreover, we derive upper limits on several decays that have not been searched for experimentally yet, such as $D^0\to e\tau$ in the charm sector, and various semileptonic decays such as $B\to \rho \mu\tau$, $B_s\to K \mu\tau$ and $B_s\to\phi\mu\tau$. Lastly, we discuss the validity of the EFT description of LHC data and the impact of loop corrections in our analysis.
\vspace{3mm}
\end{abstract}

\maketitle

\allowdisplaybreaks

\section{Introduction}\label{sec:intro}

Precision measurements of low-energy flavor observables are complementary to the effort led by the LHC in the high-energy frontier to search for physics beyond the Standard Model (SM). Flavor processes are fundamental in this quest as they are sensitive to new physics scales well beyond the reach of direct searches, providing information about the flavor structure beyond the SM. In particular, one of the most prominent types of low-energy observables are decays with Lepton Flavor Violation (LFV), which are strictly forbidden by accidental symmetries in the SM. Therefore, these are clean probes of New Physics (see Ref.~\cite{Calibbi:2017uvl} for a recent review).

The LHC is currently leading the effort in the search for new-physics particles in the high-energy frontier. 
 Besides performing direct searches for particles that might be light enough to be produced on-shell, the LHC can also be used to look for indirect effects from new particles that are too heavy to be produced in proton collisions~\cite{Farina:2016rws}. The latter strategy is based on the study of the high-energy tails of kinematic distributions, and it has been proven to be extremely useful to constrain flavor transition through LHC data on the Drell-Yan processes $pp\to \ell_i \ell_j$ and $pp\to \ell_i \nu$ at high-$p_T$, see~Refs.~\cite{Allwicher:2022gkm,Allwicher:2022mcg} and references therein. By exploiting the five quark flavors that are available in the proton, several Effective Field Theory (EFT) studies have been performed in the literature, deriving LHC bounds on specific semileptonic $d=6$ operators that were compared to low-energy constraints~\cite{Cirigliano:2012ab,Angelescu:2020uug,Fuentes-Martin:2020lea}. Although less accurate in most cases, the energy enhancement of the partonic cross sections induced by the effective operators can lead to LHC bounds that outperform the ones derived from low-energy processes for specific operators and quark/lepton flavors, as shown, for instance, for semileptonic LFV transitions in Ref.~\cite{Angelescu:2020uug} and for charm-meson decays in Ref.~\cite{Angelescu:2020uug,Fuentes-Martin:2020lea}.

\begin{figure}[t!]
\centering
\includegraphics[width=0.95\linewidth]{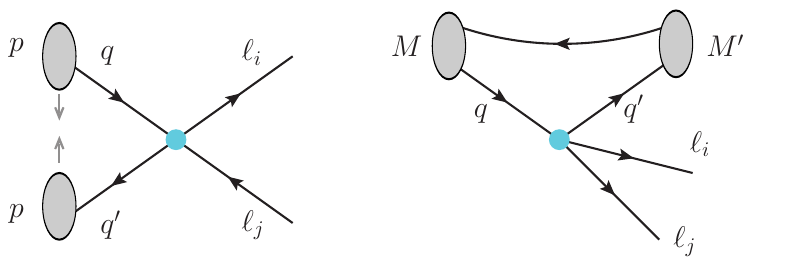}
\caption{\small \sl Illustration of the complementarity between di-lepton production at the LHC (left panel) and low-energy meson decays (right panel) to constrain semileptonic LFV interactions. }
\label{fig:illustration} 
\end{figure}

In this letter, we will demonstrate that Drell-Yan data can be used to set fully model-independent upper limits on LFV meson decays. In other words, we will derive upper limits on the  branching fractions for these processes without making any assumptions regarding the choice of effective operators. We will only assume that the EFT approach is valid to describe the LHC data~\cite{Brivio:2022pyi}. It will be fundamental for our derivation that Drell-Yan processes are inclusive, probing operators with all possible Lorentz structures and all possible combinations of the five quark flavors at tree level. This is in contrast to exclusive meson decays which are sensitive to specific quark-flavor transitions and to operators with selected Lorentz structures, depending on the quantum numbers of the mesons in the initial and final states, as depicted in Fig.~\ref{fig:illustration}.

We will illustrate the above procedure for the decays $B_{(s)}\to \ell_i\ell_j$ and $B_{(s)}\to 
M \ell_i\ell_j$, with $i\neq j$, where $M$ denotes a pseudoscalar or vector meson~\cite{Becirevic:2016zri}. We will show that current LHC data is already sufficient to provide competitive bounds for specific channels compared to the direct limits from flavor experiments, which will be further improved with the high-luminosity phase of the LHC. We will also use LHC data to constrain decays that have not yet been searched for experimentally, such as $D^0\to e\tau$ in the charm sector and various semileptonic decays of $B_{(s)}$-mesons. These results can provide a valuable guideline for low-energy experimental searches, indicating the range of branching fractions that are not yet constrained by LHC data.

The main caveat of our derivation is the validity of the EFT to describe LHC data, which must be verified in each case by comparing \textit{a posteriori} the energy of LHC events $(E)$ with the EFT scales ($\Lambda$) that can be probed for effective coefficients ($\mathcal{C}$), within the range given by perturbativity~\cite{Brivio:2022pyi}. 
If the experimental sensitivity is poor, it is only possible to probe large values of $\mathcal{C}/{\Lambda}^2$, which would either imply that $\mathcal{C}$ takes large values, or that the EFT cutoff does not satisfy the consistency condition $E\ll \Lambda$, cf.~Sec.~\ref{sec:numerical}. Another important aspect is the operator mixing induced by the Renormalization Group Evolution (RGE) of electroweak and Yukawa interactions, which spoils the tree-level correspondence displayed in Fig.~\ref{fig:illustration}. For the processes that we consider, we will show that these effects are suppressed by loop and CKM factors, thus being negligible. However, that is not the case in general, as we will briefly show for quarkonia decays at the end of Sec.~\ref{sec:numerical}.

The remainder of this letter is organized as follows: In Sec.~\ref{sec:eft}, we define our effective Lagrangian for LFV processes. In Sec.~\ref{sec:lhc}, we discuss the LFV bounds that can be derived from LHC data. In Sec.~\ref{sec:indirect}, we demonstrate that the latter can be used to constrain meson decays and we discuss the impact of one-loop operator mixing on our results. Our numerical results are presented in Sec.~\ref{sec:numerical} and we briefly conclude in Sec.~\ref{sec:conclusion}.

\section{EFT approach for LFV}\label{sec:eft}

We start by defining our framework. Since we are interested in New Physics effects arising well above the electroweak scale, we consider an effective Lagrangian invariant under the $SU(3)_c \times SU(2)_L \times U(1)_Y$ gauge symmetry, namely the SM EFT~\cite{Buchmuller:1985jz,Grzadkowski:2010es}, with operators up to dimension six,
\begin{align}
    \mathcal{L}_\mathrm{eff} = \sum_a \dfrac{\mathcal{C}_a}{\Lambda^2} \mathcal{O}_a\,,
\end{align}

\noindent where $\Lambda$ denotes the EFT cutoff. The effective coefficients are generically denoted by $\mathcal{C}_a$ and the effective operators $\mathcal{O}_a$ can be of several types. The most relevant ones for both low- and high-energy LFV searches are the semileptonic ones ($\psi^4$), which are collected in Table~\ref{tab:SMEFT-ope}. We adopt the same notation of Ref.~\cite{Jenkins:2013zja,Jenkins:2013wua,Alonso:2013hga} and denote flavor indices by Latin symbols~\footnote{The only difference with respect to Ref.~\cite{Jenkins:2013zja} is that the operator $\mathcal{O}_{qe}$ is replaced by $\mathcal{O}_{eq}$ so that the first two indices always correspond to lepton flavors.}. For Hermitian operators, we further impose that the Wilson coefficient $[\mathcal{C}]_{ijkl}$ must have $i \leq j$ to avoid redundancy in the operator basis from the relation $[\mathcal{C}]_{ijkl}=[\mathcal{C}]_{jilk}^\ast$~\cite{Allwicher:2022gkm}. For definiteness, we opt to work in the basis where down-quark Yukawas are diagonal, i.e.~the quark doublets correspond to $q=(V^\dagger u_L,\,d_L)^T$ where $V$ is the Cabibbo–Kobayashi–Maskawa (CKM) matrix.

Besides semileptonic operators, the dipole ($\psi^2 X H$) and Higgs-current operators ($\psi^2 D H^2$) can also violate lepton flavor, but these are already tightly constrained by purely leptonic processes. Moreover, such operators would induce suppressed contributions to the tails of $pp\to\ell_i\ell_j$ with respect to the semileptonic ones, since they do not feature the same scaling with the collision energy~\cite{Allwicher:2022gkm} and, therefore, will be neglected in the following.

\section{ LFV in $pp\to \ell_i\ell_j$}\label{sec:lhc}

The $q_k \bar{q}_l \to \ell_i^- \ell_j^+$ partonic cross section (with $q=u,d$) can be written as~\cite{Angelescu:2020uug}
\begin{equation}
  \label{eq:partonic-xsection}
 \hat{\sigma} (q_k \bar{q}_l \to \ell_i^- \ell_j^+) = \dfrac{\hat{s}}{144 \pi \Lambda^4} \sum_{AB} \mathcal{C}_A^{(q)\ast} \,\Omega_{AB} \,\mathcal{C}_B^{(q)}\,,
\end{equation}
where $\hat{s}$ is the partonic center-of-mass energy, and $\mathcal{C}^{(u)}$ and $\mathcal{C}^{(d)}$ are vectors of effective coefficients,
\begin{small}
\begin{align}\label{eq:hybrid_basis}
\mathcal{\vec{C}}^{\,(u)} &= \left[\mathcal{C}_{lq}^{\prime\,(1-3)}\,,\;\mathcal{C}_{lu}\,,\;\mathcal{C}_{eq}^\prime\,,\;\mathcal{C}_{eu}\,,\;\mathcal{C}_{lequ}^{\prime\,(1)} \,,\;\mathcal{C}_{lequ}^{\prime\,(1)\,\dagger}\,,\;\mathcal{C}_{lequ}^{\prime\,(3)} \,,\;\mathcal{C}_{lequ}^{\prime\,(3)\,\dagger} \right]\,,\nonumber \\[0.4em]
\mathcal{\vec{C}}^{\,(d)} &= \left[\mathcal{C}_{lq}^{(1+3)}\,,\;\mathcal{C}_{ld}\,,\;\mathcal{C}_{eq}\,,\;\mathcal{C}_{ed}\,,\;\mathcal{C}_{ledq}\,,\;\mathcal{C}_{ledq}^\dagger \,,\;0\,,\;0\right]\,,
\end{align}
\end{small}

\begin{table}[!t]
\renewcommand{\arraystretch}{2.}
\centering
\begin{tabular}{cc}
\centering
\begin{tabular}{c|c}
$\psi^4$ &   Operator\\ \hline\hline
$\mathcal{O}_{lq}^{(1)}$ & $\big{(}\bar{l} \gamma^\mu l\big{)}\big{(}\bar{q} \gamma_\mu q\big{)}$ \\ 
$\mathcal{O}_{lq}^{(3)}$ & $\big{(}\bar{l} \gamma^\mu \tau^I l\big{)}\big{(}\bar{q} \gamma_\mu \tau^I q\big{)}$\\ 
$\mathcal{O}_{lu}$ & $\big{(}\bar{l} \gamma^\mu l\big{)}\big{(}\bar{u} \gamma_\mu u\big{)}$\\ 
$\mathcal{O}_{ld}$ & $\big{(}\bar{l} \gamma^\mu l\big{)}\big{(}\bar{d} \gamma_\mu d\big{)}$\\ 
$\mathcal{O}_{eq}$ & $\big{(}\bar{e} \gamma^\mu e\big{)}\big{(}\bar{q} \gamma_\mu q\big{)}$\\ 
$\mathcal{O}_{eu}$ & $\big{(}\bar{e} \gamma^\mu e\big{)}\big{(}\bar{u} \gamma_\mu u\big{)}$\\ 
$\mathcal{O}_{ed}$ & $\big{(}\bar{e} \gamma^\mu e\big{)}\big{(}\bar{d} \gamma_\mu d\big{)}$\\ 
\end{tabular}
&
\hspace{.4cm}
\begin{tabular}{c|c}
$\psi^4$ &   Operator $+\mathrm{h.c.}$\\ \hline\hline
$\mathcal{O}_{ledq}$ & $\big{(}\bar{l}^a e\big{)}\big{(}\bar{d} q^a\big{)}$ \\ 
$\mathcal{O}_{lequ}^{(1)}$ & $\big{(}\bar{l}^a e \big{)}\varepsilon_{ab}\big{(}\bar{q}^b u\big{)}$\\ 
$\mathcal{O}_{lequ}^{(3)}$ & $\big{(}\bar{l}^a \sigma^{\mu\nu}e \big{)}\varepsilon_{ab}\big{(}\bar{q}^b \sigma_{\mu\nu}u\big{)}$\\
\end{tabular}
\end{tabular}
\vspace{0.2cm}
\caption{\small \sl Hermitian (left) and non-Hermitian (right) dimension $d=6$ semileptonic operators in the SM EFT. Quark and lepton doublets are denoted by $q$ and $l$, respectively, and the weak singlets read $u$, $d$ and $e$. $SU(2)_L$ indices are denoted by $a,b$, with $\varepsilon_{12}=-\varepsilon_{21}=+1$, and $SU(3)_c$ indices are omitted. Flavor indices are also omitted in this Table.}
\label{tab:SMEFT-ope} 
\end{table}

\noindent where {we use the shorthand notation $\mathcal{C}_{lq}^{(1\pm 3)}=\mathcal{C}_{lq}^{(1)}\pm \mathcal{C}_{lq}^{(3)}$, and flavor indices are not explicitly written, but should be understood as $\mathcal{C} \equiv {[}\mathcal{C}{]}_{ijkl}$ and $\mathcal{C}^\dagger \equiv {[}\mathcal{C}{]}_{jilk}^\ast$.~\footnote{The zeros in the $\mathcal{C}^{(d)}$ reflect the fact there are no tensor effective coefficients for the $d_k\to d_l\ell_i\ell_j$ transition at $d=6$ in the SM EFT~\cite{Alonso:2014csa}.} The primed coefficients in $\mathcal{\vec{C}}^{(u)}$ are defined as follows,
\begin{align}
\mathcal{C}_{lq}^{\prime\,(1-3)} &= V \mathcal{C}_{lq}^{(1-3)} V^\dagger\,,\\[0.45em]
\mathcal{C}_{eq}^{\prime} &= V \mathcal{C}_{eq} V^\dagger\,,\\[0.45em]
\mathcal{C}_{lequ}^{\prime\,(1)} &= V  \mathcal{C}_{lequ}^{\,(1)}\,, \\[0.45em]
\mathcal{C}_{lequ}^{\prime\,(3)} &= V  \mathcal{C}_{lequ}^{\,(3)}\,,
\end{align}

\noindent where the CKM matrix $V$ acts on quark-flavor indices. Similar expressions hold for the conjugated coefficients ($\mathcal{C}^\dagger$). 
The above redefinition involving the CKM matrix corresponds to the up-quark aligned basis, which will simplify the discussion below, since these coefficients contribute to a single quark-level transition. Lastly, the $8\times 8$ real matrix $\Omega$ takes a diagonal form $\Omega_{AB}=\Omega_A \, \delta_{AB}$ for the full partonic cross section,
\begin{equation}
  \Omega =  \mathrm{diag} \big{(}1\,,\;1\,,\;1\,,\;1\,,\;3/4\,,\;3/4\,,\;4\,,\;4 \big{)}\,.
\end{equation}

\noindent The interference terms are either suppressed by the fermion masses, being negligible at the LHC, or they vanish after integration over $\hat{t}\in (-\hat{s},0)$~\cite{Allwicher:2022gkm}. Small corrections due to the lepton masses or to the experimental cuts can also be incorporated, having a minor impact on the following discussion. Note, also, that the $\psi^2 X H$ and $\psi^2 D H^2$
operators are not included in Eq.~\eqref{eq:partonic-xsection} since their contributions are further suppressed by $\smash{v/\sqrt{\hat{s}}}$ and $v^2/\hat{s}$, respectively~\cite{Allwicher:2022gkm}.

\paragraph*{LHC data treatment} The most recent LHC searches for heavy resonances in the $pp\to e\mu,e\tau,\mu\tau$ channels~\footnote{If not stated otherwise, we will omit the lepton electric-charges, implicitly considering the sum of the two processes, e.g.~$pp\to \ell_i^\pm \ell_j^\mp$.} have been made by CMS with $140~\mathrm{fb}^{-1}$~\cite{CMS:2021tau}. The dilepton production cross section at the LHC is obtained via the convolution of the partonic cross section in Eq.~\eqref{eq:partonic-xsection} with the luminosity functions $\mathcal{L}_{q_k \bar q_l}(\hat{s})$,
\begin{equation}
\label{eq:xsection-pp-ll}
\sigma(pp\to \ell_i^-\ell_j^+) =  \sum_{k,l} \int \dfrac{\mathrm{d}\hat{s}}{s} \mathcal{L}_{q_k \bar q_l}\, \hat{\sigma}(q_k \bar q_l\to \ell_i^- \ell_j^+) \,,
\end{equation}
with
\begin{equation}
\mathcal{L}_{ q_k \bar  q_l}(\hat{s}) \equiv \int_{\hat{s}/s}^1 \dfrac{\mathrm{d}x}{x}\Big{[} f_{{q}_k} (x,\mu_F) f_{\bar{q}_l} (\frac{\hat{s}}{s x},\mu_F)+({q}_k \leftrightarrow \bar q_l)\Big{]}\,,
\end{equation}
where $f_{{q}_k}$ and $f_{\bar{q}_l}$ denote the Parton Distribution Functions (PDFs) of $q_k$ and $\bar{q}_l$, and $\mu_F$ stands for the factorization scale. 

We consider the constraints on the SM EFT Lagrangian provided in the {\tt HighPT} package~\cite{Allwicher:2022gkm,Allwicher:2022mcg}, which have been obtained through the recast of the CMS searches~\cite{CMS:2021tau} using the {\tt PDF4LHC15$\_$nnlo$\_$mc} PDF set~\cite{Butterworth:2015oua}. After minimizing the $\chi^2$-distribution describing the invariant-mass tails, we can write the resulting LHC constraints at 2$\sigma$ in the following general form,
\begin{equation}
\label{eq:Cvec-LHC_quartic}
\sum_{n\,\in\,\rm{bins}}\alpha_n\,(\mathcal{\vec{C}}^{\,\dagger} A_n \mathcal{\vec{C}}\,-1)^2 \leq 1\,,
\end{equation}
where the summation spans over all experimental dilepton invariant-mass bins, $\alpha_n$ are real coefficients and $A_n$ are Hermitian matrices determined for the $n^{\rm th}$ bin. The vector $\mathcal{\vec{C}}$ comprises all SM EFT Wilson coefficients that violate lepton flavor, defined at the scale $\mu_\mathrm{high} \approx 1~\mathrm{TeV}$. To simplify our discussion, we will focus on the limits extracted from a single mass bin at large invariant mass. 
In this case, Eq.~\eqref{eq:Cvec-LHC_quartic} can be solved and brought to a quadratic form, 
\begin{equation}
\label{eq:Cvec-LHC}
\mathcal{\vec{C}}^{\,\dagger} B \mathcal{\vec{C}} \leq 1\,,
\end{equation}
where $B$ is a block diagonal matrix, with a non-zero positive-definite block corresponding to the coefficients entering the partonic processes in Eq.~\eqref{eq:partonic-xsection}. The advantage of working with a single invariant-mass bin is that the determination of bounds for low-energy LFV decays reduces to a tractable eigenvalue problem that can be solved analytically, as will be shown in Sec.~\ref{sec:indirect}. By choosing a single interval at large invariant-mass, we will find constraints that are only slightly weaker than the ones obtained by using data from the whole invariant-mass spectrum, cf.~Sec.~\ref{sec:numerical}.

Before presenting our approach to indirectly constrain low-energy LFV observables, we note that the effective coefficients $\big{\lbrace}\mathcal{C}_{eu},\,\mathcal{C}_{lu},\,\mathcal{C}^{\prime\,(1)}_{lequ},\,\mathcal{C}^{\prime\,(3)}_{lequ},\,\mathcal{C}^{\prime\,(1-3)}_{lq}\big{\rbrace}$ with third-generation quark indices do not appear in Eq.~\eqref{eq:Cvec-LHC}, since they would correspond to partonic processes with a top quark, which are not constrained by Drell-Yan processes at tree level. Therefore, the corresponding entries of the matrix $B$ vanish for these operators by construction. Note also that the non-zero block in $B$ is itself block diagonal in the basis defined in Eq.~\eqref{eq:hybrid_basis}, with the only non-diagonal part corresponding to $\mathcal{C}_{eq}$. Indeed, $\mathcal{C}_{eq}$ is the only effective coefficient in our basis that contributes to both $\smash{u_k\bar{u}_l \to \ell_i^- \ell_j^+}$ and $\smash{d_k\bar{d}_l \to \ell_i^- \ell_j^+}$, with contributions modulated by the CKM-matrix elements and by the quark PDFs.

\section{Indirect probes of low-energy LFV decays}\label{sec:indirect}

In this Section, we demonstrate that LHC data can be used to set model-independent bounds on low-energy LFV processes. This procedure will then be applied to the most relevant LFV observables based on the $b\to d\ell_i\ell_j$ and $b\to s\ell_i\ell_j$ transitions. Other decays modes will be briefly discussed in Sec.~\ref{sec:numerical}.

\subsection*{Analytical derivation} Let us consider a given low-energy meson decay related to the quark-level transition $q_k\to q_l \ell_i \ell_j$, with $i \neq j$, and let us denote its branching fraction  by $\mathcal{O}_{\mathrm{low}}$. This observable is a quadratic form in the Low-Energy EFT (LEFT) coefficients, defined at the relevant low-energy scale $\mu=\mu_\mathrm{low}$. 
We have then to account for the RGE effects induced by QCD and QED from $\mu_\mathrm{low}$ to $\mu=\mu_\mathrm{ew}$~\cite{Jenkins:2017dyc,Crivellin:2017rmk}, for the tree-level matching to the SM EFT Lagrangian at the electroweak scale~\cite{Jenkins:2017jig}, and for the gauge and Yukawa RGE effects above the electroweak scale~\cite{Jenkins:2013zja,Jenkins:2013wua,Alonso:2013hga}. Once this is done, it is possible to write the observable $\mathcal{O}_{\mathrm{low}}$ as
 \begin{equation}
     \label{eq:olow}
     \mathcal{O}_\mathrm{low} = \vec{\mathcal{C}}^{\,\dagger} M \vec{\mathcal{C}}\,,
 \end{equation}
\noindent where the SM EFT Wilson coefficients $\vec{\mathcal{C}}$ are expressed at the scale $\mu_\mathrm{high} \approx 1$~TeV{, and $M$ is a Hermitian matrix. The RGE effects induced by QCD are numerically significant since they change the magnitude of scalar and tensor contributions to $\mathcal{O}_\mathrm{low}$. Although smaller, the RGE effects induced by the electroweak and Yukawa interactions cannot be neglected, since they can induce mixing between the operators appearing in Eq.~\eqref{eq:partonic-xsection} with those that do not contribute to these processes at tree level. In other words, the tree-level correspondence between low-energy meson decays and Drell-Yan processes in Fig.~\ref{fig:illustration} can be spoiled by RGE effects. These unwanted loop contributions have to be bounded by other means when they are numerically significant, e.g.~by using other LHC processes or electroweak/flavor observables; or simply by imposing a naive perturbativity bound, as we choose to do in the following~\footnote{Note that these effects are negligible for the processes that we consider in Sec.~\ref{sec:indirect}, thus simplifying our analysis.}.

The problem that we want to solve can be summarized as the determination of
 \begin{equation}
    \label{eq:optimization-problem-1}
    \mathcal{O}_\mathrm{max}= \max_{\vec{\mathcal{C}}} \Big{\lbrace}  \vec{\mathcal{C}}^{\,\dagger} M \vec{\mathcal{C}}\,;~\; \text{with}~\; \vec{\mathcal{C}}^{\,\dagger} B \vec{\mathcal{C}}\leq 1  \Big{\rbrace}\,,
 \end{equation}
where we remind that the matrix $B$ is defined in Eq.~\eqref{eq:Cvec-LHC}. To distinguish the operators appearing in Eq.~\eqref{eq:partonic-xsection} from the ones appearing only through RGE effects, we define the projector $P$ onto the Wilson coefficients entering 
Eq.~\eqref{eq:partonic-xsection}, i.e.~the matrix $-P$ is the projector to the kernel of $B$. With this definition, we can decompose $\smash{\vec{\mathcal{C}}= P \,\vec{\mathcal{C}} + (1-P)\,\vec{\mathcal{C}}}$ and identify the spurious operators as being those proportional to $(1-P) \,\vec{\mathcal{C}}$. The problem defined in Eq.~\eqref{eq:optimization-problem-1} can thus be decomposed as
 \begin{equation}
    \label{eq:optimization-problem-2}
    \mathcal{O}_\mathrm{max} = \mathcal{O}_\mathrm{DY}+\mathcal{O}_{\slashed{\mathrm{DY}}}\,,
 \end{equation}
where the first term is fully constrained by Drell-Yan data,
 \begin{equation}
 \label{eq:optimization-problem-3}
\mathcal{O}_\mathrm{DY} \equiv \max_{\vec{\mathcal{C}}} \Big{\lbrace} \vec{\mathcal{C}}^{\,\dagger} M_P \vec{\mathcal{C}}\,;~\; \text{with}~\; \vec{\mathcal{C}}^{\,\dagger} B \vec{\mathcal{C}}\leq 1  \Big{\rbrace}\,,
 \end{equation}
where $M_P \equiv P^\dagger  M P$ and $B=P^\dagger B P$ by construction. The second term in Eq.~\eqref{eq:optimization-problem-2} must be bounded by other means since it depends on coefficients that do not contribute to $pp\to \ell_i \ell_j$. In our case, we use a naive perturbativity constraint,
 \begin{align}
 \label{eq:non-DY}
\mathcal{O}_\slashed{\mathrm{DY}}(\Lambda) \equiv &\max_{\vec{\mathcal{C}}} \Big{\lbrace} \vec{\mathcal{C}}^{\,\dagger} (1-P) M (1-P) \vec{\mathcal{C}}\\*
&+\vec{\mathcal{C}}^{\,\dagger} P M (1-P) \vec{\mathcal{C}}\nonumber\\* &+ \vec{\mathcal{C}}^{\,\dagger} (1-P)  M P \vec{\mathcal{C}}\,;~\; \text{with}~\; ||\vec{\mathcal{C}}|| \leq 4\pi (v/\Lambda)^2 \Big{\rbrace} \,,\nonumber
\end{align}
which will depend on the EFT cutoff $\Lambda$. The norm $||\vec{\mathcal{C}}||$  is defined as the maximum norm, i.e.~the largest entry of $\vec{\mathcal{C}}$ in absolute value. The problem is now reduced to finding $\mathcal{O}_\mathrm{DY}$ and $\mathcal{O}_{\slashed{\mathrm{DY}}}$ separately, which are well-defined numbers that are \emph{necessarily bounded} for each process.

To determine $\mathcal{O}_{\mathrm{DY}}$, we first diagonalize the LHC matrix as $B=U^\dagger \hat{B} U$, where $\hat{B}$ is a diagonal matrix, and $U$ is a unitary matrix. It is convenient to define $\vec{{\zeta}}\equiv (V{\smash{\hat{B}^{1/2}}} U)\, \vec{\mathcal{C}}$, so that the LHC constraint can be simply written as $\smash{\vec{{\zeta}}^\dagger \vec{{\zeta}} \leq 1}$, and the low-energy observable takes the canonical quadratic form $\vec{{\zeta}}^\dagger\hat {N}_P \vec{{\zeta}}$ in terms of $\vec{{\zeta}}$, where $\hat{N}_P$ is diagonal. Here, the matrix $\hat{N}_P$ results from diagonalizing the Hermitian matrix $N_P\equiv{\smash{\hat{B}^{-1/2}}}U\, M_P\, U^\dagger{\smash{\hat{B}^{-1/2}}}$ via unitary rotations $N_P=V^\dagger \hat{N}_P V$. Therefore, the optimization problem in Eq.~\eqref{eq:optimization-problem-3} is now reduced to
 \begin{equation}
    \label{eq:optimization-problem-4}
\mathcal{O}_\mathrm{DY} = \max_{\vec{\zeta}} \Big{\lbrace} \vec{\zeta}^{\,\dagger} \hat{N}_P \vec{\zeta}\,;~\; \text{with}~\; \vec{\zeta}^{\,\dagger} \vec{\zeta}\leq 1  \Big{\rbrace}\,.
 \end{equation}
In other words, we have to evaluate the unit vector $\vec{\zeta}$ along the quadratic form given by $\hat{N}_P$. For simplicity, we assume that the diagonal entries of $\hat{N}_P$ are ranked in descending order. The maximal value of the quadratic form will be reached along the smallest semi-axis of the corresponding ellipsoid, which is given by the largest eigenvalue of $\hat{N}_P$, i.e.,~${\mathcal{O}_{\mathrm{DY}} = \mathrm{max}_i\lbrace {(}\hat{N}_P{)}_{ii}\rbrace={(}\hat{N}_P{)}_{11}}$, since the eigenvalues of $\hat{N}_P$ are ordered. The effective coefficient that maximized $\mathcal{O}_{\mathrm{DY}}$ is then obtained by the corresponding eigenvector,
\begin{equation}
    \label{eq:Cmax}
    \mathcal{C}^\mathrm{max}_i = \big{(}U^\dagger \hat{B}^{-1/2}V^\dagger \big{)}_{i1}\,,
\end{equation}
which provides a useful cross-check of the EFT validity, as will be discussed in Sec.~\ref{sec:numerical}.

To summarize the above discussion, we have shown that a low-energy observable $\mathcal{O}_\mathrm{low}$ can be expressed in a quadratic form in terms of the SM EFT effective coefficients that describe Drell-Yan processes, see~Eq.~\eqref{eq:xsection-pp-ll}. When using a single invariant-mass bin in the $\chi^2$-function describing LHC data, we can express the LHC constraints as a quadratic form as well, cf.~Eq.~\eqref{eq:Cvec-LHC}. Since RGE effects can introduce non-trivial mixing between operators, the quadratic form describing $\mathcal{O}_\mathrm{low}$ can be split into a quadratic form that is constrained by Drell-Yan processes ($\mathcal{O}_{\mathrm{DY}}$) and another one that is not ($\mathcal{O}_{\slashed{\mathrm{DY}}}$), cf.~Eq.~\eqref{eq:optimization-problem-2}. The latter term can be bounded, e.g.~by perturbativity or by considering other observables, if they are numerically significant, whereas the maximal value of $\mathcal{O}_{\mathrm{DY}}$ can be obtained by solving the corresponding eigenvalue problem, which has a unique solution. We reiterate that the analytical method described above produces slightly weaker upper-limits on the LFV meson decays than the ones obtained by considering binned data from the whole invariant-mass spectrum studied experimentally at the LHC, as we will provide in Sec.~\ref{sec:numerical}. Therefore, this derivation is not only a demonstration that the maximization problem is well defined, but it is also a useful cross-check for the numerical optimization considering all experimental bins.

\subsection*{Illustration: LFV $B$-meson decays} 

In this Section, we apply the above approach to LFV $B_{(s)}$-meson decays based on the $b\to d\ell_i \ell_j$ and $b\to s\ell_i \ell_j$ transitions~\cite{Becirevic:2016zri}. Several searches for these processes have been performed at the $B$-factories~\cite{BaBar:2012azg,Belle:2021rod,Belle:2022pcr} and LHCb~\cite{LHCb:2019ujz,LHCb:2022wrs} in the past years. We considered these decays as an illustration of our approach since they violate both quark and lepton flavors. For this reason, contributions from spurious operators in the second term of Eq.~\eqref{eq:optimization-problem-2} can be safely neglected since they are suppressed not only by loop factors but also CKM matrix elements. However, we note that our method can also be applied to other processes such as quarkonia decays where RGE effects are not entirely negligible, as we will briefly discuss in Sec.~\ref{sec:numerical} (see also Ref.~\cite{Calibbi:2022ddo}).

\paragraph*{EFT description} The low-energy effective Lagrangian needed to describe the $b\to q \ell_i^- \ell_j$ processes (with $q=s,d$) reads
\begin{align}
\begin{split}    
\label{eq:left-bsll}
\mathcal{L}_\mathrm{eff} \supset \dfrac{1}{v^2} \sum_{X,Y} &\Big{[}C_{V_{XY}}^{q\,ij} \big{(} \bar{\ell}_i \gamma_\mu P_X\ell_j\big{)}\big{(}\bar{q} \gamma^\mu P_Y b\big{)} \\
    &+C_{S_{XY}}^{q\,ij} \big{(} \bar{\ell}_i  P_X\ell_j\big{)}\big{(}\bar{q}  P_Y b\big{)}\Big{]}+\mathrm{h.c.}\,,
\end{split}
\end{align}
where $v=(\sqrt{2} G_F)^{-1/2}$ denotes the SM vacuum expectation value and we assume that $i<j$, as before. Notice that tensor operators are not written in this equation because they are forbidden for the $b\to q \ell_i\ell_j$ transitions, at dimension $d=6$, by the $SU(2)_L\times U(1)_Y$ gauge symmetry~\cite{Alonso:2014csa}; cf.~Table~\ref{tab:SMEFT-ope}. Only the scalar operators are renormalized by QCD, which amounts to $\smash{C_{S_{XY}}^{ij} (m_b) \simeq 1.46\times C_{S_{XY}}^{ij} (m_Z)}$~\cite{Gonzalez-Alonso:2017iyc}. 

The tree-level matching of Eq.~\eqref{eq:left-bsll} to the SM EFT at the scale $\mu=\mu_\mathrm{ew}$ reads
\begin{align}
C_{V_{LL}}^{q\,ij} &= \dfrac{v^2}{\Lambda^2}\big{[}\mathcal{C}_{lq}^{(1+3)}\big{]}_{ijq3}\,,  &C_{S_{LL}}^{q\,ij} &= 0\,,\\[0.35em]
C_{V_{LR}}^{q\,ij} &=  \dfrac{v^2}{\Lambda^2}\big{[}\mathcal{C}_{ld}\big{]}_{ijq3}\,,  &C_{S_{LR}}^{q\,ij} &=  \dfrac{v^2}{\Lambda^2}\big{[}\mathcal{C}_{ledq}\big{]}_{ji3q}^\ast\,,\\[0.35em]
C_{V_{RL}}^{q\,ij} &= \dfrac{v^2}{\Lambda^2}\big{[}\mathcal{C}_{eq}\big{]}_{ijq3}\,,  &C_{S_{RL}}^{q\,ij} &= \dfrac{v^2}{\Lambda^2}\big{[}\mathcal{C}_{ledq}\big{]}_{ijq3}\,,\\[0.35em]
C_{V_{RR}}^{q\,ij} &= \dfrac{v^2}{\Lambda^2}\big{[}\mathcal{C}_{ed}\big{]}_{ijq3}\,,  &C_{S_{RR}}^{q\,ij} &= 0\,,
\end{align}
where the Higgs current operators ($\psi^2 D H^2$) do not appear, since we consider processes that violate both lepton and quark flavors. The SM EFT operators appearing in the right-hand side of the above equations are affected by RGE from $\mu_\mathrm{ew}$ up to $\Lambda\approx 1~\mathrm{TeV}$~\cite{Jenkins:2013zja,Jenkins:2013wua,Alonso:2013hga}. QCD running will only change the magnitude of scalar coefficients, with~$\mathcal{C}_{ledq}(\mu_\mathrm{ew})\simeq 1.19\times\mathcal{C}_{ledq}(1~\mathrm{TeV})$. Instead, the electroweak and, most importantly, the Yukawa running effects will introduce nontrivial mixing between different types of operators~\cite{Feruglio:2016gvd}. For instance, keeping only the top-quark Yukawa $y_t$ contributions to the anomalous dimensions of $\mathcal{C}_{lq}^{(1)}$~\cite{Jenkins:2013wua},
\begin{align}
\label{eq:clq1-running}
\big{[}\dot{\mathcal{C}}_{lq}^{(1)}&\big{]}_{ijkl}  \stackrel{\text{yuk}}{=}  \big{[}Y_u^\dagger Y_u\big{]}_{kl}  \big{[}\mathcal{C}_{Hl}^{(1)}\big{]}_{ij} -\big{[}Y_u^\dagger\big{]}_{kv}\big{[}Y_u\big{]}_{wl} \big{[}\mathcal{C}_{lu}\big{]}_{ijvw} \nonumber \\*[0.35em]
&\hspace{-1.13em}+\dfrac{1}{2}\big{[}Y_u^\dagger Y_u \big{]}_{kv} \, \big{[}{\mathcal{C}}_{lq}^{(1)}\big{]}_{ijvl}+\dfrac{1}{2}\big{[}Y_u^\dagger Y_u \big{]}_{vl} \,\big{[}{\mathcal{C}}_{lq}^{(1)}\big{]}_{ijkv} \,,
\end{align}

\begin{figure}[t!]
\centering
\includegraphics[width=0.95\linewidth]{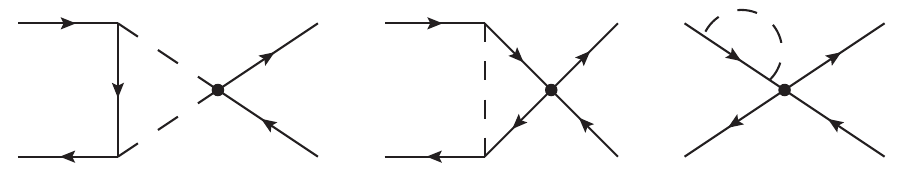}
\caption{\small \sl Example of the Yukawa-induced RGE mixing of $\psi^2 D H$ operators (left) and $\psi^4$ (middle and right) into the operator $\mathcal{O}_{lq}^{(1)}$, which contribute to the $b\to d \ell_i\ell_j$ and $b\to s \ell_i\ell_j$ transitions. The dots represent the insertion of $d=6$ operators, cf.~Eq.~\eqref{eq:clq1-running}.}
\label{fig:illustration2} 
\end{figure}

\noindent where $\dot{\mathcal{C}} \equiv 16 \pi^2 \frac{\mathrm{d} \mathcal{C}} {\mathrm{d} \log \mu}$, the up-type Yukawa matrix is defined by $Y_u= \mathrm{diag}(y_u,y_c,y_t)\cdot V \simeq \mathrm{diag}(0,0,y_t) \cdot V$, and $\mathcal{O}_{Hl}^{(1)}$ and $\mathcal{O}_{Hl}^{(3)}$ are the Higgs-current operators defined as follows~\cite{Jenkins:2013zja}
\begin{align}
\begin{split}
\big{[}\mathcal{O}_{Hl}^{(1)}\big{]}_{ij}&=\big{(}H^\dagger \overleftrightarrow{D}_\mu H\big{)}\big{(} \bar{l}_i \gamma^\mu l_j\big{)}\,,\\[0.35em]
\big{[}\mathcal{O}_{Hl}^{(3)}\big{]}_{ij}&=\big{(}H^\dagger  \overleftrightarrow{D}_\mu^I H\big{)}\big{(} \bar{l}_i\tau^I \gamma^\mu l_j\big{)}\,.
\end{split}
\end{align}

\noindent By choosing in the left-hand side of Eq.~\eqref{eq:clq1-running} the Wilson coefficient that contributes to $b\to d\ell_i\ell_j$ at tree level, we find
\begin{align}
\big{[}\dot{\mathcal{C}}_{lq}^{(1)}\big{]}_{ij13} &\stackrel{\text{yuk}}{=} y_t^2\, V_{td}^\ast V_{tb}\big{[}\mathcal{C}_{Hl}^{(1)}\big{]}_{ij}  - y_t^2\,V_{td}^\ast V_{tb} \big{[}\mathcal{C}_{lu}\big{]}_{ij33} \\[0.35em]
& +\dfrac{y_t^2}{2} V_{td}^\ast V_{tb}\big{[}\mathcal{C}_{lq}^{(1)}\big{]}_{ij33}  + \dots\,,  \nonumber
\end{align}
where the ellipsis denotes terms suppressed by small Yukawa couplings and/or CKM matrix elements.
From the above equation, we see for instance that $\smash{\mathcal{C}_{Hl}^{(1)}}$ and $\mathcal{C}_{lu}$ mixes into $\smash{\mathcal{C}_{lq}^{(1)}}$ through the top-quark Yukawa running, as illustrated in the first two diagrams of Fig.~\ref{fig:illustration2}. These are examples of operators that are not constrained by $\smash{pp \to \ell_i \ell_j}$ tails at tree level, and are thus part of $(1-P)\vec{C}$ in the derivation of \eqref{eq:optimization-problem-2}. Nonetheless, in the specific case that we consider, these contributions will be small due to the loop and CKM suppression which make $\mathcal{O}_{\slashed{\mathrm{DY}}}$ largely subdominant~\footnote{We stress once again that this is not necessarily the case for quark-flavor conserving processes such as quarkonia decays, as we shall briefly comment in Sec.~\ref{sec:numerical}.}.

In the following, we provide the expressions for LFV $B_{(s)}$-meson branching fractions in terms of the low-energy effective Lagrangian~\eqref{eq:left-bsll} and describe the hadronic inputs that we consider in our analysis. These expressions will be combined with the RGE effects described above, within a leading logarithmic approximation, to write the observables in terms of the SM EFT coefficients at the scale $\Lambda$ that is relevant for the LHC analysis.

\paragraph*{$B_{(s)}\to \ell_i \ell_j$} The branching fractions of the leptonic $B_{(s)}$-meson decays reads
\begin{small}
\begin{align}
    \label{eq:B-leptonic}
     &\mathcal{B}(P \to \ell_i^- \ell_j^+)=    \tau_{P} \, \dfrac{f_P^2 M m_{\ell}^2}{ 128 \pi  v^4} \bigg{(}1-\dfrac{m_{\ell}^2}{M^2}\bigg{)}^2 \\* 
    &\times\bigg{\lbrace}\bigg{|}C_{VA}^{q\,ij}+ \dfrac{M^2\,C_{SP}^{q\,ij}}{m_{\ell}(m_b+m_q)}\bigg{|}^2- \bigg{|}C_{AA}^{q\,ij}+ \dfrac{M^2\,C_{PP}^{q\,ij}}{m_{\ell}(m_b+m_q)}\bigg{|}^2\bigg{\rbrace} \,,\nonumber
\end{align}
\end{small}

\noindent where $P$ denotes the $B_{(s)}$ meson, with mass $M$ and lifetime $\tau_P$, and we have assumed $m_\ell \equiv  m_{\ell_j} \gg m_{\ell_i}$. The effective coefficients are defined in terms of Eq.~\eqref{eq:left-bsll} as follows,  
\begin{align}
C_{VA}^{q\,ij} &= {C_{V_{RR}}^{q\,ij}-C_{V_{RL}}^{q\,ij} - (L\leftrightarrow R)} \\[0.35em]
C_{AA}^{q\,ij} &= {C_{V_{RR}}^{q\,ij}-C_{V_{RL}}^{q\,ij} + (L\leftrightarrow R)} 
\end{align}
which should be taken at $\mu=m_b$, and similarly for the (pseudo)scalar operators through the trivial replacements $V \leftrightarrow S$ and $A \leftrightarrow P$. The $B_{(s)}$-meson decay constant is denoted
by $f_P$, which parameterizes the hadronic matrix element $\langle 0 | \bar{q} \gamma^\mu\gamma_5 b | P (p) \rangle = i f_P \,p^\mu$, where $p$ denotes the $P$-meson four-momentum. In our analysis, we consider the latest average of lattice determinations of the decay constants made by the Flavor Lattice Averaging Group (FLAG), which gives $f_B = 190.0(1.3)$~MeV and $f_{B_s} = 230.3(1.3)$~MeV~\cite{FlavourLatticeAveragingGroupFLAG:2021npn}.

\paragraph*{$B_{(s)}\to M \ell_i \ell_j$} For the $B_{(s)}\to M \ell_i \ell_j$ decays, where $M$ denotes a light pseudoscalar or vector meson, the branching fractions depend on form factors which parameterize the relevant hadronic matrix elements. In this letter, we use the general expressions provided in Ref.~\cite{Becirevic:2016zri} and we update the numerical values with the most recent determinations of the form factors, namely:
\begin{itemize}
    \item For the decays into a pseudoscalar meson, there are only two relevant form factors for our analysis, namely the scalar $(f_0)$ and vector $(f_+)$,  which have been computed for the relevant transitions by means of numerical simulations of QCD on the lattice at high-$q^2$ values and extrapolated to the whole kinematical range by using a suitable parameterization~\cite{FlavourLatticeAveragingGroupFLAG:2021npn}. For the $B\to \pi$ transition, we consider the combined fit to lattice QCD and experimental data made by FLAG~\cite{FlavourLatticeAveragingGroupFLAG:2021npn}. For $B_s\to K$, we use the FLAG average of lattice QCD form factors~\cite{FlavourLatticeAveragingGroupFLAG:2021npn}, and for $B\to K$  we use the recent combination of lattice QCD results~\cite{Bailey:2015dka,Parrott:2022rgu} made in Ref.~\cite{Becirevic:2023aov}  (see also Ref.~\cite{Parrott:2022zte}). 
    \item For the decays into a vector meson such as $M=K^\ast,\phi$, there is less information available from lattice QCD for the relevant form factors, namely $A_0$, $A_1$, $A_2$ and $V$. For these decays, we use the Light-Cone Sum-Rules results from Ref.~\cite{Bharucha:2015bzk} (see also Ref.~\cite{Gubernari:2018wyi}).
\end{itemize}

\section{Numerical results}\label{sec:numerical}

In this Section, we apply the method outlined in Sec.~\ref{sec:indirect} to the LFV decays of $B_{(s)}$-mesons detailed above. We will focus only on the 
processes leading to $\tau e$ or $\tau \mu$ pairs, since LHC bounds cannot be competitive with the very stringent low-energy limits on the decays with $\mu e$ in the final state, as already shown in Ref.~\cite{Angelescu:2020uug}. We will present our final results by solving the optimization problem numerically, considering the whole invariant-mass spectrum provided in Ref.~\cite{CMS:2021tau}. These results will be later compared to the ones from the analytical approach introduced in Sec.~\ref{sec:indirect} when a single invariant-mass bin is considered at high-energies.

In Table~\ref{tab:exp-lfv-had-etau}, we present our upper limits for the decays based on the transitions $b\to d \mu \tau$ and $b\to s \mu\tau$ (top tier), and $b\to d e \tau$ and $b\to s e\tau$ (bottom tier) obtained from a reinterpretation~\cite{Allwicher:2022gkm,Allwicher:2022mcg} of the latest CMS search with 140~$\mathrm{fb}^{-1}$~\cite{CMS:2021tau}. These limits are compared in the same table to the direct experimental limits from BaBar~\cite{BaBar:2012azg}, Belle~\cite{Belle:2021rod,Belle:2022pcr,Belle:2023jwr} and LHCb~\cite{LHCb:2019ujz,LHCb:2022wrs}. For simplicity, we sum over the decay modes with opposite-charge leptons, e.g.,~$\mathcal{B}(B\to \mu^\pm \tau^\mp)\equiv \mathcal{B}(B\to \mu^+ \tau^-)+\mathcal{B}(B\to \mu^- \tau^+)$ and similarly for the other channels~\footnote{Notice that the LFV branching fractions with opposite leptonic charges (i.e.,~$\smash{\ell_i^- \ell_j^+}$ vs.~$\smash{\ell_i^+ \ell_j^-}$) can differ from each other in scenarios such as the ones with low-energy leptoquarks~\cite{Becirevic:2016zri,Becirevic:2016oho}.}. We also present in the same table the projections for the High-Luminosity LHC (HL-LHC) phase with $3~\mathrm{ab}^{-1}$, obtained by assuming that uncertainties are statistically dominated.

From the comparison in Table~\ref{tab:exp-lfv-had-etau}, we conclude that flavor experiments perform better than our indirect LHC bounds for most transitions, as one would expect. However, there are exceptions such as the decays $B\to \pi e\tau$ and $B\to \pi \mu\tau$, for which both constraints are comparable with current data, with projections for the HL-LHC phase are more constraining than the existing limits from the $B$-factories. Another interesting example comes from decay modes that have not yet been searched for at low energies, such as the semileptonic channels $B_s\to K_S\ell_i \ell_j$, $B^0\to \rho\ell_i\ell_j$ and $B_s\to \phi\ell_i\ell_j$.  In this case, our results are the only bounds that are currently available and they may constitute a target for future experimental searches at low energies, as they indicate the range of the branching ratios that are not yet constrained by high-$p_T$ data.

\begin{table}[!t]
\renewcommand{\arraystretch}{2.1}
\centering
\resizebox{\columnwidth}{!}{\begin{tabular}{|c|cc|c|}
\hline 
Observable & LHC ($140~\mathrm{fb}^{-1}$)  & HL-LHC ($3~\mathrm{ab}^{-1}$) & Exp.~limit \\ \hline\hline
$\mathcal{B}(B^0 \to \mu^\pm \tau^\mp)$ & $8 \times 10^{-4}$   & $1.7 \times 10^{-4}$  & \cellcolor{blue!07}$1.4\times 10^{-5}$  \\
$\mathcal{B}(B^+ \to \pi^+ \mu^\pm \tau^\mp)$ & $1.1 \times 10^{-4}$   & \cellcolor{blue!03}$2 \times 10^{-5}$  & $9.4\times 10^{-5}$  \\
$\mathcal{B}(B_s \to K_S^0 \mu^\pm \tau^\mp)$ &  \cellcolor{blue!07}$4 \times 10^{-5}$   &  \cellcolor{blue!03}$8 \times 10^{-6}$           & --\\ 
$\mathcal{B}(B^0 \to \rho \mu^\pm \tau^\mp)$ &  \cellcolor{blue!07}$7 \times 10^{-5}$   &  \cellcolor{blue!03}$1.5 \times 10^{-5}$         & -- \\ \hline
$\mathcal{B}(B_s \to \mu^\pm \tau^\mp)$ &  $8 \times 10^{-3}$ & $1.7 \times 10^{-3}$  & \cellcolor{blue!06}$4.2\times 10^{-5}$ \\ 
$\mathcal{B}(B^+ \to K^+ \mu^\pm \tau^\mp)$ &  $9 \times 10^{-4}$ &  $1.9 \times 10^{-4}$  & \cellcolor{blue!06}$3.9\times 10^{-5}$  \\
$\mathcal{B}(B^0 \to K^{\ast\,0} \mu^\pm \tau^\mp)$ &  $4 \times 10^{-4}$   & $1.0 \times 10^{-4}$     & \cellcolor{blue!07}$2.2\times 10^{-5}$ \\
$\mathcal{B}(B_s \to \phi \mu^\pm \tau^\mp)$ &  \cellcolor{blue!07}$5 \times 10^{-4}$   & \cellcolor{blue!03}$1.0 \times 10^{-4}$      & -- \\ \hline\hline
$\mathcal{B}(B^0 \to e^\pm \tau^\mp)$ &  $1.7 \times 10^{-3}$  & $4 \times 10^{-4}$  & \cellcolor{blue!07} $2.1\times 10^{-5}$ \\
$\mathcal{B}(B^+ \to \pi^+ e^\pm \tau^\mp)$ &  $2 \times 10^{-4}$   & \cellcolor{blue!03} $5 \times 10^{-5}$  &  $9.8\times 10^{-5}$ \\
$\mathcal{B}(B_s \to K_S e^\pm \tau^\mp)$ & \cellcolor{blue!07}$8 \times 10^{-5}$   &  \cellcolor{blue!03}$1.7 \times 10^{-5}$          & -- \\ 
$\mathcal{B}(B^0 \to \rho e^\pm \tau^\mp)$ &  \cellcolor{blue!07}$1.4 \times 10^{-4}$   &   \cellcolor{blue!03}$3 \times 10^{-5}$        & -- \\ \hline
$\mathcal{B}(B_s \to e^\pm \tau^\mp)$ &  $1.8 \times 10^{-2}$  & $4 \times 10^{-3}$  &\cellcolor{blue!07} $7.3\times 10^{-4}$ \\ 
$\mathcal{B}(B^+ \to K^+ e^\pm \tau^\mp)$ & $2 \times 10^{-3}$ &  $4 \times 10^{-4}$  &  \cellcolor{blue!07}$3.9\times 10^{-5}$ \\
$\mathcal{B}(B^0 \to K^{\ast\,0} e^\pm \tau^\mp)$ &   \cellcolor{blue!07}$1.1 \times 10^{-3}$    & \cellcolor{blue!03} $2 \times 10^{-4}$     & -- \\
$\mathcal{B}(B_s \to \phi e^\pm \tau^\mp)$ &  \cellcolor{blue!07}$1.2 \times 10^{-3}$    &  \cellcolor{blue!03}$2 \times 10^{-4}$       & --\\ \hline
\end{tabular}}
\caption{\small \sl Upper limits on the $B_{(s)}$-meson branching fractions based on the transitions $b\to d\tau \mu$ and $b\to s\tau \mu$ (top tier), and $b\to d\tau e$ and $b\to s\tau e$ (bottom tier), obtained by using our approach with current LHC data (140~$\mathrm{fb}^{-1}$) at 95$\%$ CL, as well as our projections for HL-LHC (3~$\mathrm{ab}^{-1}$). The last column corresponds to the direct experimental limit obtained experimentally by BaBar~\cite{BaBar:2012azg}, Belle~\cite{Belle:2021rod,Belle:2022pcr,Belle:2023jwr} and LHCb~\cite{LHCb:2019ujz,LHCb:2022wrs}. The highlighted cells correspond to the most stringent limits with current (dark blue) or future (light blue) data.}
\label{tab:exp-lfv-had-etau} 
\end{table}

From Table~\ref{tab:exp-lfv-had-etau}, we also find that our constraints on semileptonic decays such as $B\to \pi \tau \ell$ and $B\to \rho\tau \ell$, with $\ell=e,\mu$, are one order of magnitude stronger than the ones on the leptonic channel $B_d\to \tau \ell$. To understand this difference, one should note that purely leptonic decays are very sensitive to (pseudo)scalar operators since they can induce contributions that are larger by a factor of $m_{B_s}^2/m_\tau^2$ than the vector ones, cf.~Eq.~\eqref{eq:B-leptonic}. This is not the case for semileptonic decays, nor for the high-$p_T$ processes, which receive comparable contributions from the different Lorentz structures. This is illustrated in Fig.~\ref{fig:plots-Bpimutau} for the decays based on the transition $b\to d \mu\tau$, by considering the scalar effective coefficient $\mathcal{C}_{ledq}$ and the vector $\smash{\mathcal{C}_{lq}^{(1)}}$ with fixed flavor indices.

\begin{figure*}[t!]
\centering
\includegraphics[width=0.33\linewidth]{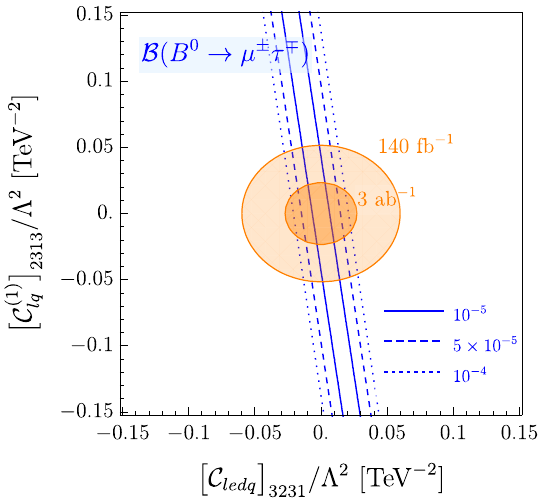}~\includegraphics[width=0.33\linewidth]{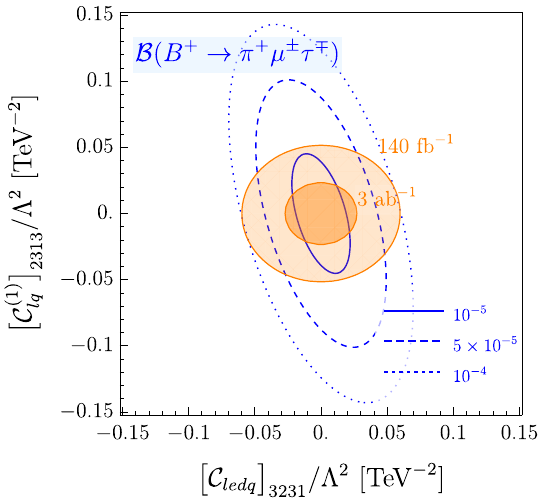}~\includegraphics[width=0.33\linewidth]{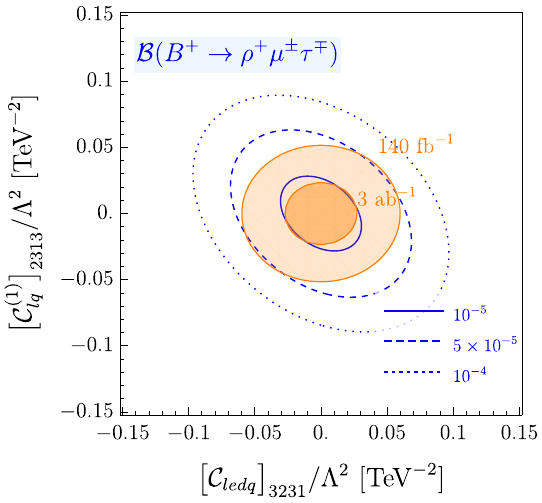}
\caption{\small \sl The contour lines for $\mathcal{B}(B^0\to \mu^\pm \tau^\mp)$ (left), $\mathcal{B}(B^+\to \pi^+ \mu^\pm \tau^\mp)$ (center) and $\mathcal{B}(B^+\to \rho^+ \mu^\pm \tau^\mp)$ (right) are depicted in the plane $\smash{\big{[}\mathcal{C}_{ledq}^{(1)}\big{]}_{3231}}/\Lambda^2$ vs.~$\smash{\big{[}\mathcal{C}_{lq}^{(1)}\big{]}_{2313}}/\Lambda^2$ by the blue lines, with the other effective coefficients set to zero. The 95$\%$~CL constraints derived from current LHC data (140~fb$^{-1}$) and their projection to HL-LHC (3~ab$^{-1}$) are depicted by the dark- and light-orange regions.  }
\label{fig:plots-Bpimutau} 
\end{figure*}

Lastly, we comment on the differences between the numerical maximization of $\mathcal{O}_\mathrm{low}$ taking into account the whole dilepton invariant-mass spectrum for the LHC constraints~\cite{Allwicher:2022gkm,Allwicher:2022mcg}, as reported in Table~\ref{tab:exp-lfv-had-etau}, with the simplified problem obtained in Eq.~\eqref{eq:optimization-problem-1} by considering a single high invariant-mass bin. To this purpose, we computed the matrix $B$ defined in Eq.~\eqref{eq:Cvec-LHC} for the intervals $m_{e\tau}^2 \in (1550,3000)$~GeV and $m_{\mu\tau}^2\in(1400,3000)$~GeV for the $pp\to e\tau$ and $pp\to\mu\tau$ channels, respectively, and we performed the analytical derivation from Sec.~\ref{sec:indirect}. With this approach, we are able to derive bounds that are slightly weaker than the full limits colected in Table~\ref{tab:exp-lfv-had-etau}, with deviations of at most $\approx 40\%$. Therefore, this comparison corroborates the validity of the analytical derivation of Sec.~\ref{sec:indirect}, which is a valuable cross-check of the numerical optimization.

\paragraph*{On the EFT validity } Before discussing the application of the method to other processes, we briefly comment on the validity of the EFT description of LHC data~\cite{Brivio:2022pyi}. For the EFT to be valid, the EFT cutoff ($\Lambda$) must be much larger than the energy scale ($E$) of the relevant LHC events. Since we are constraining combinations of the type $\mathcal{C}/\Lambda^2$, it is clear that for a search of given sensitivity, our limits will only apply for $\mathcal{C}$ above a certain value. The maximal EFT cutoff $\Lambda_\mathrm{max}$ for which our results are applicable can be estimated by requiring that the $2\to 2$ amplitude does not exceed the $4 \pi$ perturbativity bound~\cite{Farina:2016rws}. Assuming that the effective operators are generated at tree level, $\Lambda_\mathrm{max}$ is given by
\begin{equation}
    \label{eq:max-cutoff}
    \Lambda_\mathrm{max} = \dfrac{\sqrt{4\pi} }{\sqrt{\mathcal{C}_\mathrm{max}/\Lambda^2}}\,,
\end{equation}

\noindent which must satisfy $\Lambda < \Lambda_\mathrm{max}$, where $\mathcal{C}_\mathrm{max}$ denotes the largest effective coefficient that maximizes the observable in Eq.~\eqref{eq:olow} (see also Eq.~\eqref{eq:Cmax}). For the constraints derived in Table~\ref{tab:exp-lfv-had-etau}, we find that the maximal cutoff $\Lambda_\mathrm{max}$ is sufficiently above $E$ for all processes. For instance, we obtain that
\begin{align}
\Lambda_\mathrm{max} &\approx 16~\mathrm{TeV}\,, &&(B\to \pi \mu \tau)\,\\[0.25em]
\Lambda_\mathrm{max} &\approx 13~\mathrm{TeV}\,, &&(B\to \pi e \tau)\,\\[0.25em]
\Lambda_\mathrm{max} &\approx 12~\mathrm{TeV}\,, &&(B\to K \mu \tau)\,,\\[0.25em]
\Lambda_\mathrm{max} &\approx 8~\mathrm{TeV}\,, &&(B\to K e \tau)\,,
\end{align}

\noindent with similar values for the other processes depending on the same low-energy transitions. The largest effective coefficient for the leptonic $P\to \ell \tau$ decays and for the semileptonic  $P\to P^\prime \ell \tau$ is the scalar $\smash{\mathcal{C}_{ledq}}$, with appropriate flavor indices, whereas the semileptonic $P\to V \ell \tau$ processes are maximized by vector operators, in agreement with the findings in Fig.~\ref{fig:plots-Bpimutau}.

Finally, in order to determine which energy bins are the most relevant for our bounds, we plot in the right panel of Fig.~\ref{fig:plots-clipping} our upper limits on $\mathcal{B}(B\to\pi\mu\tau)$ as a function of  a fixed energy scale $\Lambda_\mathrm{cut}$ above which events with higher di-lepton invariant masses are neglected. The last bins are indeed the most important ones, as expected from the energy enhancement of the cross section. However, we see that our limits remain similar even if a smaller value of $\Lambda_\mathrm{cut}\approx 1$~TeV is taken. In the left panel of the same plot, we compare the current and projected constraints on the representative effective coefficient ${[\mathcal{C}_{ledq}]_{3231}}$ with the requirement that $\Lambda_\mathrm{cut}<\Lambda_\mathrm{max}$ [cf.~Eq.~\eqref{eq:max-cutoff}], which is satisfied for all values of $\Lambda_\mathrm{cut}$ in our analysis.

\paragraph*{Implications to other processes} The same method described above can be used to constrain other types of LFV meson decays. For instance, $D$-meson decays based on the $c\to u\ell_i\ell_j$ transition can be induced by an EFT Lagrangian similar to Eq.~\eqref{eq:left-bsll}, but with different flavor indices, and with the inclusion of tensor operators that are allowed by $SU(2)_L\times U(1)_Y$ gauge invariance in this case. The discussion of the RGE-induced operators that are not constrained by Drell-Yan processes is very similar to the case of $B$-meson decays, with contributions that are entirely negligible due to the CKM and loop suppressions. We find that our indirect limit for the $D^0 \to \mu e$ decay is not useful, as it is 
about two orders of magnitude weaker than the direct experimental limit, namely $\mathcal{B}(D^0\to e^\pm\mu^\mp)^\mathrm{exp}<1.6\times 10^{-8}$~\cite{ParticleDataGroup:2022pth}. However, we note that there are no experimental limits available for the analogous decay $D^0\to e \tau$, which has a very narrow phase space. In this case, we obtain the following indirect limit with our approach,
\begin{align}
\mathcal{B}(D^0\to e^\pm \tau^\mp) \leq 2\times 10^{-7}\,,\quad (95\%~\mathrm{CL})\,,
\end{align}
where we used $f_D=212.0(0.7)$~MeV for the $D$-meson decay constant~\cite{FlavourLatticeAveragingGroupFLAG:2021npn}. For this process, we find that the maximal cutoff defined in Eq.~\eqref{eq:Cmax} is given by $\Lambda_\mathrm{max} \approx 18~\mathrm{TeV}$. Note, in particular, that the corresponding decays for $\tau \mu$ transition do not exist, since $m_{\tau}-m_\mu <m_{D^0}<m_{\tau}+m_\mu$ implies that both $D^0\to \tau\mu$ and $\tau\to D^0\mu$ are kinematically forbidden.

\begin{figure*}[t!]
\centering
\includegraphics[width=.5\linewidth]{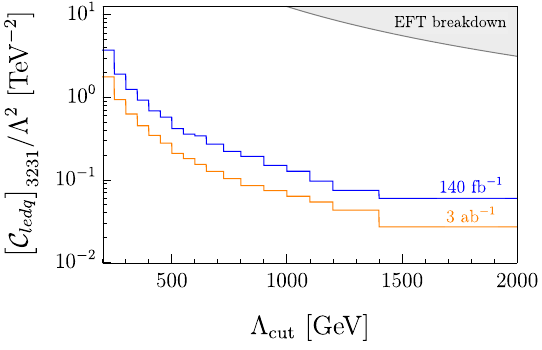}~\includegraphics[width=.5\linewidth]{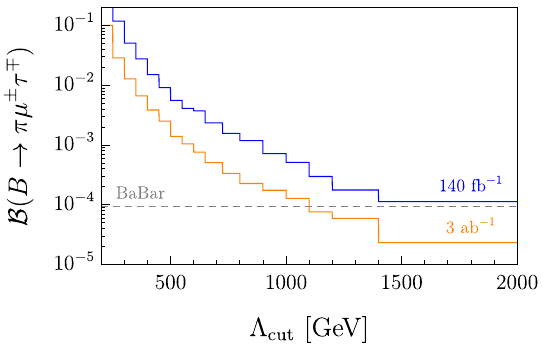}
\caption{\small \sl Left panel: upper limits on ${[\mathcal{C}_{ledq}]_{3231}}/\Lambda^2$ as a function of the energy scale $\Lambda_\mathrm{cut}$ above which the events are neglected, by using current LHC data (blue line) and our projection from HL-LHC (orange line). The shaded gray region corresponds to $\Lambda_\mathrm{cut}> \Lambda_\mathrm{max}$, cf.~Eq.~\eqref{eq:max-cutoff}.
Right panel: Upper limits on $\mathcal{B}(B^+ \to \pi^+ \mu^\pm \tau^\mp)$ as a function of the energy scale $\Lambda_\mathrm{cut}$ by considering all SM EFT operators. The current experimental limit from BaBar is depicted by the gray dashed line~\cite{BaBar:2012azg}.}
\label{fig:plots-clipping} 
\end{figure*}

In principle, the same approach could also be used to constrain LFV decays of quarkonia~\cite{Abada:2015zea}. However, important differences arise in this case since quarkonia are states without open flavor. By using our LHC bounds to $95\%$~CL and by focusing at first on the semileptonic operators from Table~\ref{tab:SMEFT-ope} that contribute at tree level to these decays, we obtain the very stringent limits from LHC data,
\begin{align}
\label{eq:quarkonia-tree}
\begin{split}
\mathcal{B}(\Upsilon \to \mu^\pm\tau^\mp)_\mathrm{tree} &\leq  3 \times 10^{-9} \,,\\[0.4em]
\mathcal{B}(\Upsilon \to e^\pm\tau^\mp)_\mathrm{tree}   &\leq  8 \times 10^{-9} \,,\\[0.4em]
\mathcal{B}(\Upsilon \to e^\pm\mu^\mp)_\mathrm{tree}    &\leq  2 \times 10^{-9} \,,
\end{split}
\end{align}
where we use the decay constant $f_\Upsilon=649(31)~\mathrm{MeV}$ computed in Ref.~\cite{Colquhoun:2014ica}. These limits correspond to $\mathcal{O}_\mathrm{DY}$ in Eq.~\eqref{eq:optimization-problem-1}, which are particularly stringent due to the large $\Upsilon$ lifetime stemming from the fact that,
differently from $B$-mesons, quarkonia can decay through electromagnetic and strong interactions~\cite{Angelescu:2020uug}. These values are orders of magnitude more constraining than the searches performed at the $B$-factories, which set upper limits in the $10^{-7}$--$10^{-6}$ range \cite{Belle:2022cce}. However, these results are misleading as they only refer to the tree-level contributions from semileptonic operators in Eq.~\eqref{eq:optimization-problem-2} (i.e.,~$\mathcal{O}_\mathrm{DY}$). We stress that  there are additional contributions that are not constrained by Drell-Yan  processes (i.e., included in  $\smash{\mathcal{O}_\slashed{\mathrm{DY}}}$).
The first contribution of this kind arises from the Higgs-current operators $\smash{\mathcal{O}_{Hl}^{(1)}}$ and $\smash{\mathcal{O}_{Hl}^{(3)}}$  which induce quarkonia decays at tree level, but which are not efficiently constrained by high-$p_T$ Drell-Yan data since they are not energy enhanced~\cite{Allwicher:2022gkm,Allwicher:2022mcg}. The second effect is induced by the RGE mixing of operators that are not constrained by Drell-Yan data into the ones needed at low energies, which is not CKM suppressed for this transition~\cite{Calibbi:2022ddo}. For instance, keeping only the $\mathcal{C}_{lu}$ effective coefficient, with couplings to the top-quark at $\mu \approx 1~\mathrm{TeV}$, and neglecting the other contributions, we obtain
\begin{equation}
\label{eq:quarkonia-loop}
\begin{small}
\hspace{-0.3em}\mathcal{B}(\Upsilon \to \mu^\pm\tau^\mp)_\mathrm{loop} \simeq    1.5 \times 10^{-9} \,\dfrac{\big{|} \big{[}\mathcal{C}_{lu}\big{]}_{2333}/(4\pi)\big{|}^2}{(\Lambda/~1~\mathrm{TeV})^4}+\dots
\end{small}
\end{equation}
where we have considered both gauge and Yukawa running effects, within a leading logarithmic approximation, and $\Lambda$ is fixed to $1~\mathrm{TeV}$ in the logarithm. By setting this effective coefficient to the naive perturbativity limit $\mathcal{C}\approx 4\pi$, we find that the loop contribution becomes comparable to the tree-level ones in Eq.~\eqref{eq:quarkonia-tree}. Therefore, one should consider additional constraints for the $\mathcal{O}_{\slashed{\mathrm{DY}}}$ terms in order to provide a meaningful bound on the full observable.

In order to constrain the tree-level contributions to $\mathcal{B}(\Upsilon\to\mu\tau)$ induced by  $\smash{\mathcal{C}_{Hl}^{(1,3)}}$,  we can consider $Z$-boson LFV decays, which should satisfy the LHC limit $\mathcal{B}(Z\to\mu^\pm\tau^\mp)<6.5 \times 10^{-6}$ at 95$\%$ CL~\cite{ATLAS:2014vur}. To estimate the impact of this constraint, we first consider $\smash{\mathcal{C}_{Hl}^{(1,3)}}$ and set the other coefficients to be zero, relating the two decay modes at tree level, 
\begin{equation}
\hspace{-0.65em}\Gamma(\Upsilon\to\mu^\pm\tau^\mp)_\mathrm{tree}   \stackrel{\mathcal{C}_{Hl}^{(1,3)}\neq 0}{\simeq}  g_{Vb}^2\dfrac{f_\Upsilon^2 m_\Upsilon^3 }{v^2 m_Z^3 }\, \Gamma(Z\to \mu^\pm\tau^\mp)\,,
\end{equation}
where $g_{Vb}=-1/2+2/3\, \sin^2 \theta_W$, $\theta_W$ denotes the Weinberg angle and, for simplicity, we have neglected the terms suppressed by $m_\tau^2/m_\Upsilon^2$ and $m_\Upsilon^2/m_Z^2$. The above expression then gives
\begin{equation}
\hspace{-0.8em}\mathcal{B}(\Upsilon\to\mu^\pm\tau^\mp)_\mathrm{tree} \stackrel{\mathcal{C}_{Hl}^{(1,3)}\neq 0}{\simeq} {2.8\times 10^{-10}}\, \dfrac{\mathcal{B}(Z\to\mu^\pm\tau^\mp) }{(6.5 \times 10^{-6})}\,,
\end{equation}
which confirms that these contributions to $\Upsilon$ decays are negligible when $Z$-pole constraints are taken into account. The $Z$-pole observables can also be used to constrain the semileptonic operators with the top quark, since they contribute to $Z\to \ell_i\ell_j$ at one loop~\cite{Feruglio:2016gvd,Calibbi:2022ddo,Coy:2019rfr}. For instance, by only keeping the $[\mathcal{C}_{lu}]_{2333}$ coefficient, we find that the $Z$-pole observables allows us to derive the upper bound, 
\begin{equation}
\dfrac{|[\mathcal{C}_{lu}]_{2333}|}{\Lambda^2} \lesssim {1.5}~\mathrm{TeV}^{-2}\,,  
\end{equation}
where we have set $\Lambda=1~\mathrm{TeV}$ in the logarithm. This bound indeed is more constraining than the perturbativity one in Eq.~\eqref{eq:quarkonia-loop}~\footnote{We notice that even more stringent bounds can be used by using the purely leptonic decays $\tau\to \mu\mu\mu$~\cite{Calibbi:2022ddo}}. This rough exercise indicates that it is possible to constraint the $\mathcal{O}_\slashed{\mathrm{DY}}$ terms in Eq.~\eqref{eq:optimization-problem-2} for $\Upsilon$ decays by combining Drell-Yan data with the $Z$-pole observables. However, it is clear that the precise assessment of the combined constraints on these decay modes would require a dedicated analysis beyond tree-level, which accounts for all relevant operators and the potentially large one-loop RGE effects.

In conclusion, Drell-Yan processes are very efficient to constrain the semileptonic operators contributing at tree-level to $\Upsilon\to\ell_i\ell_j$, but there are additional contributions to these decays at tree- and loop-level which are not negligible and that therefore need to be precisely assessed. Similar conclusions apply to LFV decays of the $\phi$ and $J/\psi$ mesons. The way out consists of a combined fit of the LHC constraints with the experimental limits on the processes $Z\to \ell_i \ell_j$ and $\ell_i \to \ell_j \ell_k \ell_k$ with $i>j\geq k$, which would allow us to apply the same procedure described in Sec.~\ref{sec:indirect} to these processes, as suggested by the above exercise with selected operators. Such an extended analysis lies beyond the scope of the present letter. 

\section{Summary and outlook}\label{sec:conclusion}

In this letter, we revisited the constraints on LFV meson decays by using LHC data on the Drell-Yan processes $pp\to\ell_i\ell_j$ (with $i\neq j$) at high-$p_T$. By relying on the EFT approach, we have shown that it is possible to derive upper bounds on such processes in a model-independent way by marginalizing over the complete set of SM EFT operators compatible with such LHC bounds. In particular, this optimization problem can be reduced to an eigenvalue problem that can be solved analytically, provided a single bin at large di-lepton invariant-mass is considered for the LHC data.

As an illustration of this approach, we have derived indirect limits on $B_{(s)}$-meson LFV decays based on the transitions $b\to d \tau \ell$ and $b\to s \tau \ell$, with $\ell=e,\mu$. We find that our upper limits on the decays $B\to \pi e \tau$ and $B\to \pi \mu\tau$ are already competitive with the current limits from the $B$-factories and that the expected sensitivity at the HL-LHC will supersede them. Furthermore, we have derived constraints of the order of $10^{-4}$ on other LFV decays that have not been searched yet experimentally, such as $B\to \rho \tau \ell$, $B_s\to K \tau \ell$ and $B_s\to \phi \tau \ell$, among others. Lastly, using the same approach, we show that the branching fractions of the charmed-meson decay $D^0\to e \tau$ must be smaller than around $10^{-7}$, for which there is not a direct experimental search yet.

The main caveat of our analysis is the validity of the EFT description of LHC data. We have estimated the maximal EFT cutoff $\Lambda_\mathrm{max}$ that can be probed in a perturbative scenario by analyzing the effective coefficients that maximize the low-energy observable. We find that our bounds can probe values of $\Lambda$ up to $\Lambda_\mathrm{max} \approx 10~\mathrm{TeV}$ for $B$-decays and $\Lambda_\mathrm{max} \approx 20~\mathrm{TeV}$ for $D$-meson decays. Our limits are therefore applicable to EFT scenarios with an EFT cutoff $\Lambda$ that satisfies $E\ll \Lambda < \Lambda_\mathrm{max}$, where $E$ denotes the energy of LHC events and $\Lambda_\mathrm{max}$ is defined in Eq.~\eqref{eq:max-cutoff}. For scenarios with light mediators, our constraints have to be reassessed by accounting for the propagation of the new degrees of freedom at the LHC, see e.g.~Ref.~\cite{Allwicher:2022gkm,Allwicher:2022mcg}.

Finally, we stress that our method can also be applied to other LFV processes such as the decays of $\phi$, $J/\psi$ and $\Upsilon$. Since LHC constraints on the operators that contributed at tree level to these processes are very stringent, one should carefully assess the RGE-induced contributions of the operators that are not constrained by Drell-Yan processes at tree level. We have argued that it is possible to extend our analysis to also constrain these contributions, e.g.~by combining the Drell-Yan constraints at high-$p_T$ with the $Z$-pole observables.
We believe that our study offers a very convincing illustration of the complementarity of high-$p_T$ and low-energy searches for LFV through the SM EFT, and we hope that our results will invite experimental collaborations to further improve the sensitivity on both Drell-Yan processes and LFV meson decays.

\

\section*{Acknowledgments}

 This project has received support from the European Union’s Horizon 2020 research and innovation programme under the Marie Skłodowska-Curie grant agreement No~860881-HIDDeN. I.~P.~receives funding from “P2IO LabEx (ANR-10-LABX-0038)” in the framework “Investissements d’Avenir” (ANR-11-IDEX-0003-01) managed by
the Agence Nationale de la Recherche (ANR), France. The work of D.A.F.~is supported
by the US Department of Energy under grant DOE-DE-SC0010008.


\end{document}